\documentclass[prb,twocolumn,superscriptaddress,floats,showpacs]{revtex4}

\usepackage[dvips]{graphicx}
\usepackage{amsmath}
\usepackage{booktabs}
\usepackage{dcolumn}

\newcommand{\Ang}{$\text{\AA}$}
\newcommand{\Angrz}{$\text{\AA}^{-1}$}

\newcommand{\tzg}{t$_\text{2g}$}
\newcommand{\dxy}{d$_\text{xy}$}
\newcommand{\dxz}{d$_\text{xz}$}
\newcommand{\dyz}{d$_\text{yz}$}
\newcommand{\muB}{$\mu_\text{B}$}
\newcommand{\ruo}{RuO$_\text{6}$}
\newcommand{\sro}{Sr$_\text{2}$RuO$_\text{4}$}
\newcommand{\csro}{Ca$_\text{2-x}$Sr$_\text{x}$RuO$_\text{4}$}

\newcommand{\srdzs}{Sr$_\text{3}$Ru$_\text{2}$O$_\text{7}$}

\newcommand{\cro}{Ca$_\text{2}$RuO$_\text{4}$}
\newcommand{\csrx}{Ca$_\text{2-x}$Sr$_\text{x}$RuO$_\text{4}$}

\newcommand{\csrea}{Ca$_\text{1.8}$Sr$_\text{0.2}$RuO$_\text{4}$}
\newcommand{\csref}{Ca$_\text{1.5}$Sr$_\text{0.5}$RuO$_\text{4}$}
\newcommand{\csreda}{Ca$_\text{1.38}$Sr$_\text{0.62}$RuO$_\text{4}$}

\newcommand{\kf}{k$_\text{f}$}
\newcommand{\Qfm}{Q$^\text{FM}$}

\newcommand{\astar}{$a^\star$}
\newcommand{\bstar}{$b^\star$}
\newcommand{\cstar}{$c^\star$}
\newcommand{\Qab}{Q$^{IC}_{\alpha\beta}$}
\newcommand{\Qice}{Q$^{IC}_{1}$}
\newcommand{\Qicz}{Q$^{IC}_{2}$}

\begin{document}

\advance\vsize by 2 cm

\title{Magnetic excitations in the metallic single-layer Ruthenates \csrx ~
studied by inelastic neutron scattering}

\author{P. Steffens}
\email{steffens@ill.eu} \affiliation{II. Physikalisches Institut,
Universit\"{a}t zu K\"{o}ln, Z\"{u}lpicher Str.\ 77, D-50937
K\"{o}ln, Germany} \affiliation{Institut Laue Langevin, 6 Rue
Jules Horowitz BP 156, F-38042 Grenoble CEDEX 9, France}

\author{O. Friedt}
\affiliation{II. Physikalisches Institut, Universit\"{a}t zu K\"{o}ln, Z\"{u}lpicher Str.\ 77, D-50937 K\"{o}ln,
Germany}

\author{Y. Sidis}
\affiliation{Laboratoire L\'eon Brillouin, C.E.A./C.N.R.S., F-91191 Gif-sur-Yvette CEDEX, France}

\author{P. Link}
\thanks{Spektrometer PANDA, Institut f\"ur Festk\"orperphysik, TU Dresden}
\affiliation{Forschungsneutronenquelle Heinz Maier-Leibnitz (FRM-II), TU M\"unchen, Lichtenbergstr.\ 1, 85747 Garching,
Germany}

\author{J. Kulda}
\affiliation{Institut Laue Langevin, 6 Rue Jules Horowitz BP 156, F-38042 Grenoble CEDEX 9, France}

\author{K. Schmalzl}
\affiliation{IFF, Forschungszentrum J\"ulich GmbH, JCNS at ILL, F-38042 Grenoble Cedex 9, France}

\author{S. Nakatsuji}
\affiliation{Institute for Solid State Physics, University of Tokyo, Kashiwa, Chiba 277-8581, Japan}

\author{M. Braden}
\email{braden@ph2.uni-koeln.de} \affiliation{II. Physikalisches Institut, Universit\"{a}t zu K\"{o}ln, Z\"{u}lpicher
Str.\ 77, D-50937 K\"{o}ln, Germany}

\date{\today}


\begin{abstract}

By inelastic neutron scattering, we have analyzed the magnetic
correlations in the paramagnetic metallic region of the series
\csrx , $0.2 \le x \le 0.62$. We find different contributions that
correspond to 2D ferromagnetic fluctuations and to fluctuations at
incommensurate wave vectors \Qice=(0.11,0,0), \Qicz=(0.26,0,0) and
\Qab=(0.3,0.3,0). These components constitute the measured
response as function of the Sr-concentration x, of the magnetic
field and of the temperature. A generic model is applicable to
metallic \csrx\ close to the Mott transition, in spite of their
strongly varying physical properties. The amplitude,
characteristic energy and width of the incommensurate components
vary only little as function of x, but the ferromagnetic component
depends sensitively on concentration, temperature and magnetic
field. While ferromagnetic fluctuations are very strong in
\csreda\ with a low characteristic energy of 0.2~meV at T=1.5~K,
they are strongly suppressed in \csrea, but reappear upon the
application of a magnetic field and form a magnon mode above the
metamagnetic transition. The inelastic neutron scattering results
document how the competition between ferromagnetic and
incommensurate antiferromagnetic instabilities governs the physics
of this system.

\end{abstract}

\maketitle

\section{Introduction.}

The family of layered Ruthenates is in the focus of interest since the discovery of superconductivity in \sro\
\cite{maeno94,mackenzie03} whose unconventional nature is still under debate. The series \csrx\ which arises from \sro\
by substitution of Sr by Ca exhibits a variety of exciting phenomena on its own. Though the replacement of Sr by Ca
does not change the number of charge carriers, the electronic and magnetic behavior is closely coupled to slight
structural changes and varies considerably as function of the Sr-content x \cite{nakatsuji00prl,nakatsuji00prb}. \cro\
is a Mott insulator and antiferromagnetically ordered below 110~K \cite{cao97,nakatsuji97}, while for x$>$0.18, the
ground state is metallic. The strongly enhanced values of the magnetic susceptibility indicate that the system is close
to a ferromagnetic instability around x=0.5 \cite{nakatsuji00prl}. \csrx -samples with x$\sim$0.5 exhibit a remarkably
large value of the Sommerfeld coefficient of the specific heat in the range of heavy-fermion compounds. Moreover, for
0.2$<$x$<$0.5 a metamagnetic transition from a state with low susceptibility, reminiscent of antiferromagnetic
correlations, to a state with high magnetic polarization is observed \cite{nakatsuji03,nakatsuji99jlt}. The
metamagnetic transition is also observed in the closely related bilayer Ruthenate \srdzs\ \cite{perry01}, where the
discussion about quantum-critical behavior has generated further interest. The high concentration dependence of the
ground state in \csrx \ points to a complex interplay of distinct magnetic correlations. A precise characterization of
the magnetic excitations in the layered ruthenates seems interesting due to several reasons: First, ferromagnetic
fluctuations are considered to play an important role in the superconducting pairing in \sro. As ferromagnetic
fluctuations are difficult to study in this material, insight can be more easily gained from the \csrx \ compounds. In
addition, the complex magnetism in \csrx \ arrises from the interplay between structural and orbital degrees of
freedom, which is relevant in many transition-metal oxides. An orbital-selective Mott transition has been proposed to
explain the fascinating magnetism in \csrx \cite{anisimov02}.

Before we discuss the experimental findings, we briefly introduce the general concepts to describe magnetic
fluctuations in itinerant magnets and we summarize previous studies on magnetic correlations in layered ruthenates
introducing the magnetic contributions to the magnetic scattering in \csrx \ by the aid of an intensity map measured
for x=0.2. We then discuss in separate sections the three antiferromagnetic and the ferromagnetic components of the
excitation spectrum in \csrx \ with $0.2 \le x \le 0.62$, and finally we analyze the effect of magnetic fields on these
correlations.

\section{Neutron scattering on magnetic fluctuations in itinerant magnets}

The magnetic properties of a material are characterized by its wave-vector and frequency dependent magnetic
susceptibility, $\chi(\textbf{Q},\omega)$, which contains the entire information about the static, $\omega=0$, as well
as about the dynamic phenomena. Magnetic inelastic neutron scattering (INS) measures the imaginary part of the
susceptibility $\chi''(\textbf{Q},\omega)$ as function of wave vector and of frequency (energy transfer $\hbar\omega$),
whereas macroscopic methods only access the case Q=0 and (compared to INS) $\omega\rightarrow 0$. More precisely, the
INS cross section is given by \cite{lovesey}:
\begin{equation}
\frac{d^2\sigma}{d\Omega d\omega} \propto
\frac{F^2(\textbf{Q})}{1-\exp(-\frac{\hbar\omega}{k_BT})}\cdot \chi''(\textbf{Q},\omega)
\label{eqsigma}
\end{equation}
where $F(\textbf{Q})$ is the magnetic form factor. The imaginary part, directly experimentally accessible by this
formula, is connected to the real part of the susceptibility, $\chi'(\textbf{Q},\omega)$, via the Kramers-Kronig
relation.

The spin dynamics in itinerant paramagnetic systems like the layered ruthenates has the character of fluctuations,
i.~e.\ correlations limited in space and in time, around the paramagnetic ground state. $\chi''$ is non-zero everywhere
or in large portions of {\bf Q},$\omega$ space, and it is the structure of this excitation continuum that contains the
information about the magnetic interactions.

For an itinerant compound one may analytically calculate the susceptibility $\chi''$ via the Lindhard function, i.e. by
a sum over all possible excitations of electrons from occupied into empty states. Correlation effects can be modelled
by taking into account an additional interaction parameter (see for instance Ref.\ \onlinecite{moriya}) yielding an
enhancement of the generalized susceptibility. This calculation, however, is in many materials not feasible due to the
insufficient knowledge about the electronic band structure. In such cases it is still possible to deduce expressions
for $\chi''$ that are valid near the ferromagnetic or antiferromagnetic instabilities, i.~e.\ for critical
fluctuations. Expanding the inverse susceptibility as $\chi_q^{-1}(1-i\omega/\Gamma_q)$ (see Refs.\
\onlinecite{moriya,lonzarich85}), the imaginary part of the susceptibility is conveniently written in the form
\begin{equation}
\chi''(q,\omega) = \chi_q \cdot \frac{\omega\Gamma_q}{\omega^2+\Gamma_q^2}
\label{eqrelaxor}
\end{equation}
At different $q$, the frequency spectrum has thus qualitatively the same form, which frequently is referred to as
''(single) relaxor'' in the literature, and which has its maximum at $\omega=\Gamma_q$.

$\chi_q$ is expressed as $\chi_q = \frac{\chi}{1+\xi^2(q-q_o)^2}$. The form of $\Gamma_q$ depends on the nature of the
magnetic instability, i.~e.\ if it is ferromagnetic (propagation vector $q_0$=0) or antiferromagnetic (all
$q_0$$\neq$0, i.~e.\ possibly incommensurate). In the ferromagnetic case, one obtains
\begin{equation}
\Gamma_q = \Gamma_0 \cdot \xi q \cdot (1+\xi^2q^2)
\label{eqferro}
\end{equation}
while in the antiferromagnetic case
\begin{equation}
\Gamma_q = \Gamma_0 \cdot (1+\xi^2(q-q_0)^2)
\label{eqanti}
\end{equation}
The parameters $\Gamma_0$ and $\xi$ depend on the microscopic details and on how close the system is to the magnetic
phase transition. From the phenomenological point of view, they can be used to parameterize the magnetic fluctuations
in {\bf Q},$\omega$ space in a convenient and remarkably simple way. $\xi$ defines the length scale, which may be
regarded as a correlation length, and $\Gamma_0$ defines the energy scale.

The ferromagnetic case differs from the antiferromagnetic one by the additional factor $q$ in $\Gamma_q$, which causes
$\Gamma_q$ to vanish for $q\rightarrow 0$. In spite of its continuum character, the shape of $\chi''(q,\omega)$
resembles that of a dispersive excitation. The maxima in constant energy scans lie on a curve initially linear in $q$
with slope $\xi\Gamma_0$. Due to its vague resemblance to a magnon in the ordered state, such an excitation is often
called a paramagnon.

The antiferromagnetic fluctuation yields a qualitatively different intensity distribution : It has a maximum at
($q_0$,$\Gamma_0$), whose energy width is determined by $\Gamma_0$, and the $q$-width (FWHM) is 1.7$\cdot\xi^{-1}$ at
$\omega=\Gamma_0$. At higher energies, the excitation is broader in $q$, but always peaked at $q_0$; there is thus no
dispersive feature.

An illustration of how the  intensity is distributed in both cases will be given in the context of the discussion of
the experimental results later on, see section VI.

\section{Samples and experimental aspects.}

Due to the limitations by neutron flux and by signal strength, large samples are necessary to study magnetic
excitations in layered ruthenates by INS. The experiments have been performed on single crystals of \csrx\ with
concentrations x=0.2 and 0.62, that have been obtained by a floating zone method at the University of Kyoto and had
masses of 1.9~g and 1.8~g respectively.

Different triple axis spectrometers have been used: 4F1, 4F2, 1T and 2T at the LLB, Saclay, IN14, IN12, IN22 and IN20
at the ILL, Grenoble, and Panda at FRM-II, Garching. On the thermal neutron spectrometers 1T, 2T and IN22 we used a
fixed final neutron wave vector \kf=2.662~\Angrz\ and graphite filters. On IN20 we have used the newly available
Flatcone multi-detector with \kf=3~\Angrz\ in order to obtain a mapping of the intensity distribution \cite{flatcone}.
On the cold neutron spectrometers (unless otherwise stated) \kf\ has been set to 1.5~\Angrz\ and Beryllium filters were
used to suppress higher orders. In all cases, focusing monochromators and analyzers have been employed. Magnetic fields
up to 10~T have been applied using vertical cryomagnets.

\begin{figure}[t]
\begin{center}
\includegraphics*[width=.92\columnwidth]{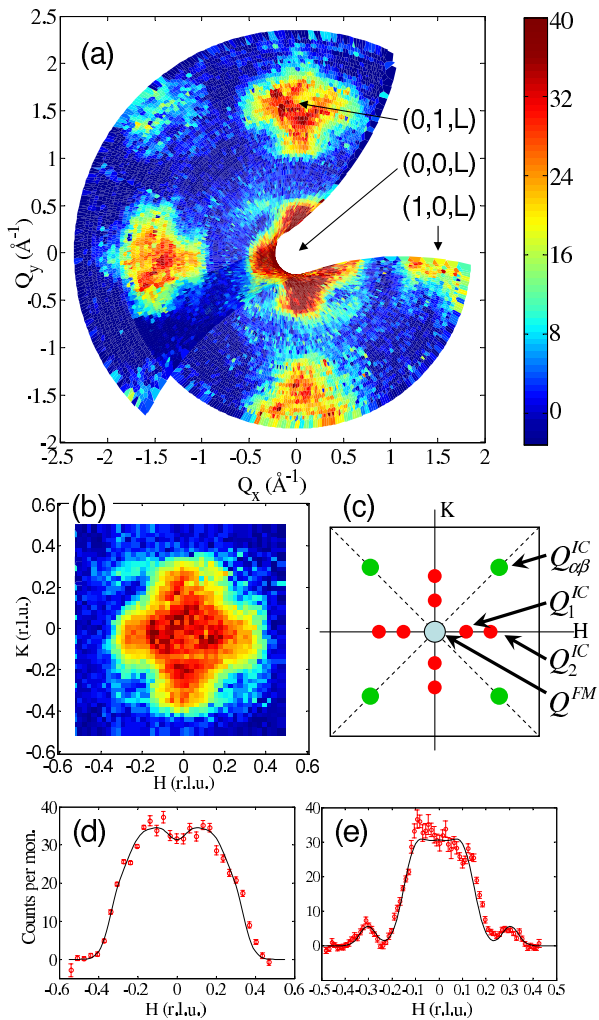}
\end{center}
\caption{(Color online) Magnetic scattering in \csrea\ at T = 2~K
and $\Delta$E = 4 meV. (a): Intensity map over several Brillouin
zones at a constant value of L=1.4. A smooth background has been
subtracted. Data was taken on IN20 using the flat-cone setup. (b):
Magnetic scattering in one Brillouin zone, obtained from the data
in (a) by projecting to the range H,K=[-0.5,0.5], thereby
averaging the scattering from different Brillouin zones (after
correction for the Ru magnetic form factor). (c): Sketch of the
(2D) Brillouin zone and the positions of the different signals.
(d): Intensity along a line Q=(H,0,1.4), obtained by integration
of the two-dimensional data set in (b). (e): The same for a
diagonal cut through the Brillouin zone through (0,0,1.4). The
data in all parts of the figure are normalized to a monitor count
rate that corresponds to approximately 1 minute counting time per
point. \label{fig1}}
\end{figure}

In order to access different regions in reciprocal space, the samples have been mounted either in [100]-[010]
orientation, i.~e.\ with \astar\ and \bstar\ in the scattering plane, or in an [100]-[001] orientation that gives
access to momentum transfers along \cstar. Here and in the following the notation is based on a unit cell that
corresponds to that of \sro, i.~e.\ a=b=3.76~\Ang, and c=12.65~\Ang\ for x=0.62 and c=12.55~\Ang\ for x=0.2. This
notation neglects the structural distortions -- a rotation of the \ruo-octahedra around the vertical axis and an
additional tilt in the case of x=0.2 \cite{friedt01} -- that have to be described in a larger unit cell
($\sqrt{2}a\cdot\sqrt{2}a\cdot 2c$). Samples were well characterized by neutron diffraction; in particular, by
regarding characteristic superstructure reflections of the structural distortions, we can state that the \csrea\ sample
was twinned with approximately equal amounts of both twins. This compound is orthorhombic with a small splitting of the
in-plane lattice constants due to the tilt distortion. Furthermore, we verified that the structure corresponds to the
so-called D-Pbca phase and not to the L-Pbca phase, which is in close proximity in the phase diagram \cite{friedt01}.
In D-Pbca, the octahedra of next-nearest layers rotate in opposite phase yielding a doubling of the c lattice
parameter.

All measurements were carried out with unpolarized neutron beams. Phonon scattering should not occur in our
experiments, as no optical phonons exist in the analyzed  energy and $q$-range \cite{braden-phon}, and the
contamination by acoustic phonons is avoided when working around Q=(1,0,0). There is no Bragg scattering at this
Q-point because of the symmetry of the crystal structure (including the distortions); but as argued below (1,0,0) can
nevertheless be considered as a magnetic zone center. The measured signals are consistent to the previous results on
magnetic excitations in the Ruthenates (for instance Refs.\ \onlinecite{friedt04,braden02sro,sidis99}), and studies of
the Q- and temperature dependence further corroborate the magnetic character. Using the lattice dynamical model
described in Ref. \onlinecite{braden-phon} we may calculate the distribution of the dynamic structure factor of an
acoustic phonon branch around Q=(2,0,0) and fold it with the experimental resolution in order to describe a phonon
scan. Using the same folding for the magnetic data it has been possible to assign absolute susceptibility units to our
data.

\section{Overview on magnetic excitations in the Ruthenates \label{secmagex}}

Magnetic excitations have so far been studied by INS in several layered Ruthenates. The most detailed description is at
present available for \sro\ (Refs.\ \onlinecite{sidis99,servant00,braden02sro,servant02,braden04ani,nagata04}).
Furthermore, the bilayer compound \srdzs\ has been investigated (Refs.\ \onlinecite{capogna03,stone06,ramos07}).The
excitations in these materials have the character of fluctuations that are both relatively broad in {\bf Q} and
$\omega$. In \sro, these fluctuations reside at {\bf Q}=(0.3,0.3,0), i.~e.\ at incommensurate wave vectors on the
diagonal of the Brillouin zone. In \srdzs, excitations at two inequivalent wave vectors have been identified that are
on the \astar\ (respectively \bstar) axis of the Brillouin zone. In \sro, no dependence on the L-component of {\bf
Q}=(H,K,L) has been found, apart the smooth decrease towards high Q that is governed by the Ruthenium magnetic form
factor. This shows that there is no relevant magnetic correlation between moments in different RuO$_2$ layers. The same
applies to \srdzs\ with the exception that there is a strong magnetic interaction between the two layers of one
double-block.

It is thus sufficient to regard the two in-plane dimensions and to neglect the L-component of Q. The 2D Brillouin zone
is quadratic, and all Q with integer H,K are zone centers, i.~e.\ any excitation at such Q, in particular at (1,0,0) or
(0,0,L), exhibits a ferromagnetic character.

In \sro, the magnetic excitation at \Qab =(0.3,0.3,L) is very well understood  on the basis of the underlying
electronic band structure\cite{mazin97}; the Fermi surface consists of three sheets, and \Qab\ is the nesting vector
connecting the so-called $\alpha$ and $\beta$ sheets which arise from the Ruthenium 4d$_{xz}$ and 4d$_{yz}$
orbitals\cite{mazin97,bergemann03}. In \srdzs, the Fermi surface is far more complex\cite{singh01,tamai08}, so it is
less simple to identify which parts of it give the relevant contribution.

In the single layer Ruthenates, however, there are other wave vectors at which magnetic correlations have been
observed: in \csrx\ with x=0.62 (Ref.\ \onlinecite{friedt04}) large contributions at the incommensurate wave vectors
(0.22,0,0) and equivalent ones have been observed; in \csrea\ (Ref.\ \onlinecite{steffens07}), similar excitations have
been seen, and even a separation in two contributions at Q$^{ic}_1$=(0.12,0,0) and Q$^{ic}_2$=(0.27,0,0) could be
resolved. In all these respects, the Ca-doped materials \csrx\ are thus fundamentally different from \sro.

Finally, ferromagnetic correlations have been identified in \csrea\ at elevated temperatures or upon application of a
magnetic field \cite{steffens07}. To what extent ferromagnetic correlations play a role also in \sro\ and its
superconductivity, remains so far an open question.

\bigskip

An overview of the distribution of scattered intensity is obtained from the data in Figure \ref{fig1} taken for \csrea,
which covers a wide region of a plane in reciprocal space. The data have been obtained using the Flatcone
multi-detector option of the IN20 spectrometer\cite{flatcone} with the sample oriented in the a,b-plane. By tilting the
sample and by placing the detector array in an inclined position out of the horizontal plane just above the direct
beam, a map of scattering vectors with a constant finite vertical component of L=1.4 is obtained. The accessible
horizontal components of Q (H and K) are small in this configuration, which is convenient for the study of magnetic
scattering due to the form factor.

In the intensity map in Figure \ref{fig1}(a) one clearly sees the magnetic intensity centered at the points with
integer H and K, in accordance to the above discussed two-dimensional character of the magnetic fluctuations. Due to
the magnetic form factor, the intensity is weaker, the higher the modulus of Q, i.~e.\ in the outer regions of the map.

The sketch in Figure \ref{fig1}(c) summarizes all the different positions of the signals that have been discussed in
the beginning of this section, see also Ref. \onlinecite{steffens07}. The cross-like intensity pattern which dominates
in the intensity map, in particular when reducing it to one (2D) Brillouin zone, see Fig.\ \ref{fig1}(b), originates
from the superposition of intensity stemming from the signals at \Qice\ and \Qicz. The resolution, effectively
broadened also by the averaging over several zones, is not sufficient to separate these contributions in the map, as it
is well possible in the data previously collected at lower energy transfer and with different experimental geometry
\cite{steffens07}. The intensity at \Qab\ is relatively weak compared to the one near the zone center,  and therefore
this \Qab -signal is only weakly visible in the colour plot. When integrating the two-dimensional data set over a
stripe of about 0.1~\Angrz width that runs along the diagonal of the Brillouin zone, these signals are clearly
distinguishable (Figure \ref{fig1}(e)).

\section{Incommensurate fluctuations}


\begin{figure}
\begin{center}
\includegraphics*[width=.95\columnwidth]{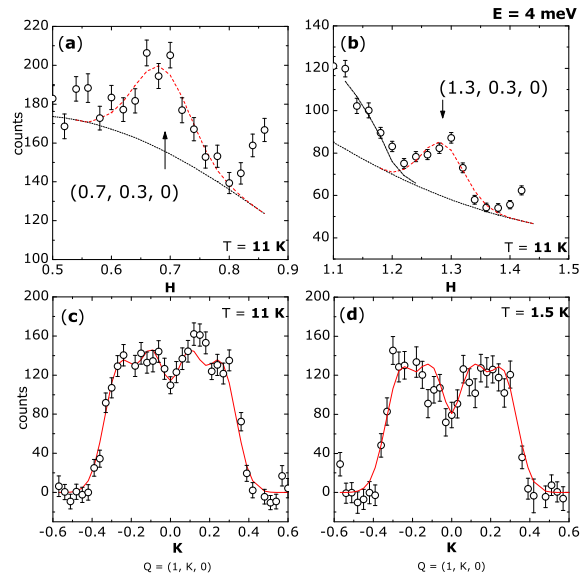}
\end{center}
\caption{(Color online) Incommensurate signals in \csreda. (a) and
(b): diagonal scans across \Qab, the thin dotted line is the
scattering angle dependent background. (c) and (d): transverse
scans (background subtracted) across (1,0,0), crossing the
positions \Qice\ and \Qicz\ at two temperatures. Scans were taken
with an energy transfer of 4~meV on the 1T spectrometer.
\label{icfig1}}
\end{figure}


In this section we discuss the different contributions to the incommensurate scattering which exhibit an
antiferromagnetic character. The observation of peaks at Q=(0.7,0.3,0) and (1.3,0.3,0) in \csreda, see Fig.\ 2(a) and
(b), proves that the signal which arises from nesting of the $\alpha$ and $\beta$ bands in \sro\ and which has been
observed in \csrea~ as well, see Fig.\ 1 and Ref.\ \onlinecite{steffens07}, is also present for x=0.62. Fitting the
position of these peaks on different equivalent Q-vectors yields H,K=0.301$\pm$0.005, i.e. the same values as in the
case of pure \sro \cite{braden02sro}. This signal thus indicates that the nesting of the $\alpha$ and $\beta$ band
remains intact despite all changes in the crystal structure and in the Fermi surfaces. First-principles
calculations\cite{ko07}  come to the conclusion that apart the back-folding effect the $\alpha$ and $\beta$ sheets of
the Fermi surface are only little affected by the rotational distortion of the structure \cite{fang01,fang04}. The
invariance of this nesting signal also suggests that the filling of these bands with respect to the $\gamma$-band does
not significantly change from the values in \sro. In a simple model that assumes a rigid filling of the band structure,
we calculate that an increase of the occupation number $n(\alpha)+n(\beta)$ by about 0.1 electron would already shift
the nesting peak more than 0.02 in H and K along the diagonal of the Brillouin zone -- by far more than the maximum
shift consistent with the experimental error bars. Even though the details of the real band structure might slightly
change this estimation, it imposes a very low boundary of the shift of electrons among the orbitals and is thus in
clear contradiction to the proposed scenario of an orbital selective Mott transition\cite{anisimov02} which requires
integer filling of the bands.

Fig.\ 2(c) and (d) show scans along the \bstar\ axis across {\bf Q}=(1,0,0) for x=0.62. The shape of the signal
resembles closely that in \csrea\ \cite{steffens07}. It exhibits steep edges at about H=$\pm$0.35 and has, apart a
minimum in the center, a broad and flat plateau in between. These features are observed in scans at a number of
different energy transfers and temperatures. A satisfactory fit can thus \emph{not} be performed with a single
symmetric (Gaussian) contribution of whatever width, but requires at least two contributions on both sides. A fit using
symmetric Gaussian peaks yields the positions $q_1 = (1, 0.10\pm 0.01, 0)$ and $q_2 = (1, 0.26\pm 0.01, 0)$, which are
nearly the same values as those reported for \csrea \cite{steffens07}.

The obtained description of the data is fully satisfying, for \csreda\ as well as for \csrea; furthermore, the
description is consistent for both cases. The large width and the significant overlap of the single contributions do
not allow us to resolve if there is even more structure intrinsic to these fluctuations or not. At least an additional
ferromagnetic component, which is discussed in detail in the next section, still plays some role at this energy and is
the likely origin of the less well pronounced minimum at (1,0,0) and the slightly lower $q_1$ in comparison to \csrea.

\subsubsection*{Relation between incommensurate signals and the
Fermi-surface }

It appears most interesting to associate the different incommensurate scattering contributions with nested parts of the
Fermi surface, similar to the case of the excitations at \Qab\ in \sro . The analysis for \csrx \ is, however, much
more difficult because the structural distortions and the large unit cell render the Fermi surface very complex, and
the present knowledge about the Fermi-surface in \csrx \ is insufficient to clearly assign the origin of the \Qice \
and \Qicz \ peaks.

There seems to be an overall consensus concerning the electronic structure of the two end members, \sro \ and \cro . In
\sro \ the four electrons are equally distributed amongst the three \tzg -bands yielding a filling of 2/3 for the three
orbitals \cite{fang01,fang04}. For this electronic arrangement the van-Hove singularity near q=(0,0.5,0) in the
$\gamma$-band associated with the \dxy -orbitals is situated  only slightly above the Fermi-energy. On the other side
of the phase diagram, the Mott-state in \cro \ is associated with orbital ordering driven through the strong structural
changes. The pronounced flattening of the octahedron in \cro \cite{braden98} results in a full occupation of the \dxy
-orbitals and in half-filled \dxz - and \dyz -orbital states undergoing the Mott transition
\cite{fang01,fang04,liebsch07,pavarini10}. Qualitatively this picture of orbital order seems to be valid for all \csro
\ undergoing the metal-insulator transition, i.e. for x$<$0.2. The electronic structure of the metallic samples at
slightly larger Sr content studied here remains matter of controversy both in experiment and in theory. Anisimov et al.
\cite{anisimov02} proposed an orbital-selctive Mott transition in the idea that correlations should first result in a
localization of the \dxz - and \dyz -states, because the corresponding bands exhibit a smaller band width in pure \sro
. The rotation of the octahedra around the c-axis, however, are found to strongly influence the $\gamma$-band
associated with the \dxy -orbitals due to the variation of the hopping \cite{fang01,fang04}. The \dxy -band essentially
flattens with the rotation and slightly shifts to lower energies whereas \dxz - and \dyz-bands are only little affected
\cite{fang01,fang04,ko07,liebsch07}. The Fermi-surface associated with the \dxy -band undergoes a topological
transition, as the van-Hove singularity shifts from above to below the Fermi level, so that the character of this band
becomes hole-like. The electronic band-structure calculations hence suggest that the $\gamma$-band associated with the
\dxy -orbitals is much closer to localization than the other bands in disagreement with the initial idea of an orbital
selective Mott transition in the \dxz -\dyz -orbitals \cite{anisimov02}.

There is strong experimental evidence that the anomalous properties of \csro\ with x$\sim$0.5 are mainly associated
with the $\gamma$-band and with the \dxy -orbitals. A spin-density study on \csref \ directly proves that the high
magnetic polarization in these phases arises from the \dxy -orbitals \cite{gukasov02} which are hybridized with
in-plane oxygen orbitals. Optical spectroscopy also identifies the \dxy -band renormalization  as the origin of the
heavy-mass behavior in \csref \cite{lee06}. Furthermore, several ARPES measurements confirm the essential modification
of the $\gamma$-band induced by the rotation of the octahedra \cite{wang04,shimoyamada09,neupane09}. Wang et al.
\cite{wang04} report a hole-like $\gamma$-band for x=0.5 in agreement with the theory and magnetoresistance
oscillations \cite{balicas05}. However, for even smaller Sr content, in the doping range where the tilt distortion is
also present, x=0.2,  Neupane et al. \cite{neupane09} report the complete disappearance of the $\gamma$-Fermi surface
sheet, whereas Shimoyamada et al. \cite{shimoyamada09} find that all bands cross the Fermi-level for this
concentration. All three ARPES studies \cite{wang04,shimoyamada09,neupane09} clearly document that the \dxy -states are
strongly renormalized. The $\gamma$-band width around the wave vectors associated with the van-Hove singularity,
(0.5,0,0), is remarkably small; it amounts to only a few meV \cite{shimoyamada09}, similar to the values reported for
\srdzs\ \cite{tamai08}. The extreme reduction of the $\gamma$-band width results in a high electronic density of states
at the Fermi level which manifests itself in the heavy-mass states \cite{nakatsuji97,nakatsuji03}  as well as in the
remarkably strong thermal expansion anomalies in \csrx\ for 0.2$<$x$<$0.6 \cite{kriener05}.

Besides the signals at \Qab\ all magnetic correlations observed in our INS experiments should be attributed to the
$\gamma$-band. The amplitude of the incommensurate signal at \Qice\ and  \Qicz\ in \csreda \cite{friedt04} is three
times larger than that of the incommensurate nesting signal in pure \sro \cite{braden02sro}. The rotational distortion
of the structure should have a relatively small effect on the \dxz- and \dyz -bands, so it appears very unlikely that
the magnetic signal associated with these bands changes so drastically between \sro\ and \csreda. This conclusion is
corroborated by the fact that the \Qab -signal, which can unambiguously be attributed to these bands, remains constant
between x=2 and x=0.2. As there is little change in the \Qice\ and \Qicz -signals for x=0.2 and 0.62, it appears
reasonable to assume that for x=0.2 this incommensurate scattering does not arise from the \dxz- and \dyz -bands
neither. Although the ferromagnetic fluctuations originate from the $\gamma$ band and the high density of states at the
Fermi level, it is still possible that some sections of the $\gamma$ surface are at the same time responsible for the
signals at \Qice\ and \Qicz \cite{ko07}. In a very simple approach, the hole-like Fermi surface for x=0.5
\cite{friedt04,wang04,liebsch07} may explain a peak in the Lindhard susceptibility associated with electron transfer
from (0.5,+$\delta$,0) to (0.5,-$\delta$,0) and thereby an incommensurate signal on the tetragonal axes, but for any
quantitative statement the knowledge about the electronic structure in \csref\ is clearly insufficient.

\subsubsection*{Contribution of the spin fluctuations to the
electronic specific heat} Like other excitations, the incommensurate magnetic fluctuations contribute to the specific
heat. Their contribution to the coefficient of the electronic specific heat at low temperature can be approximated
as\cite{moriyasph}
\begin{equation}
\gamma_{sf} = \frac{\pi k_B^2}{\hbar} \sum_q \frac{1}{\Gamma(q)}
\label{sph}
\end{equation}
where the sum is over the whole Brillouin zone. Performing a very simple estimation (taking $\Gamma$ constant in an
area defined by the widths of the peaks) with the fitted parameters for \csrea\ yields 160~mJ/molK$^2$, which is
reasonably near the directly measured value\cite{nakatsuji03} of 175~mJ/mol~K$^2$ and proves that the remarkably high
electronic specific heat can be understood in terms of the magnetic fluctuations. For \csreda, in which the coefficient
of the electronic specific heat has the even higher value of 250~mJ/molK$^2$, an analogous estimation has already been
performed\cite{friedt04} yielding perfect agreement, $\gamma_\text{sf}$=250~mJ/mol~K$^2$.

In view of the similarity of the incommensurate fluctuation spectrum of both Sr concentrations, it appears reasonable
to assume that their contribution to $\gamma_\text{sf}$ is approximately equal. The main difference between \csreda\
and \csrea\ is the presence of the ferromagnetic component; the different value of the specific heat coefficient gives
therefore a rough estimate for the contribution of the ferromagnetic fluctuations to the specific heat.

\subsubsection*{Model for the incommensurate fluctuations in \csrea\ }
As the full band structure based analysis of the magnetic excitations seems impossible at the moment, we develop a
model based on the general expressions derived  in section II. The antiferromagnetic correlations can be described more
easily in the case of \csrea, as there is no ferromagnetic part to be taken into account at low temperature.

Using equations (\ref{eqrelaxor}) and (\ref{eqanti}) for fluctuations close to a general incommensurate
antiferromagnetic instability, one obtains, in a global fit to the available data including the convolution with the
experimental resolution, a very good description of the incommensurate magnetic excitations at \Qice\ and \Qicz\ in
\csrea. To account for the two inequivalent wave vectors, each of them is assumed to contribute according to these
equations, and the contributions of these and the symmetrically equivalent positions in the Brillouin zone are simply
summed up. As the widths are large, this implies a quite significant overlap of these signals, that prohibits the
separation of the the individual contributions. $\Gamma_0$ and $\xi$ are thus set equal for both parts. It turns out
that $\xi$ can be taken as isotropic in the plane. Note, that we may neglect the orthorhombic distortion as the
tetragonal [100] and [010] directions remain equivalent in the orthorhombic lattice. Finally, we do not take into
account any anisotropy of the fluctuations in spin space -- in general, though, one would expect that in an anisotropic
system only one component of the fluctuations diverges when approaching the magnetic instability. The geometric effect
when measuring in different orientations indicates that these fluctuations are mainly polarized in the plane, i.~e.\
$\chi''_{ab}$ is significantly larger than $\chi''_c$ in contrast to the finding for the nesting signal at \Qab \ in
pure \sro \cite{braden04ani}.

The parameters that we use to describe the spin fluctuations in \csrea \ at low temperature are $q_1=0.12\pm0.01$,
$q_2=0.27\pm0.01$, $\Gamma_0=2.7\pm0.2\,\text{meV}$ and $\xi=5.9\pm0.3\,\text{\AA}$. The latter value corresponds to
less than twice the lattice spacing and shows that these correlations still exhibit a very short length scale, even
shorter than that of the fluctuations at \Qab\ in \sro\cite{sidis99,servant02}.

\begin{figure}
\begin{center}
\includegraphics*[width=.99\columnwidth]{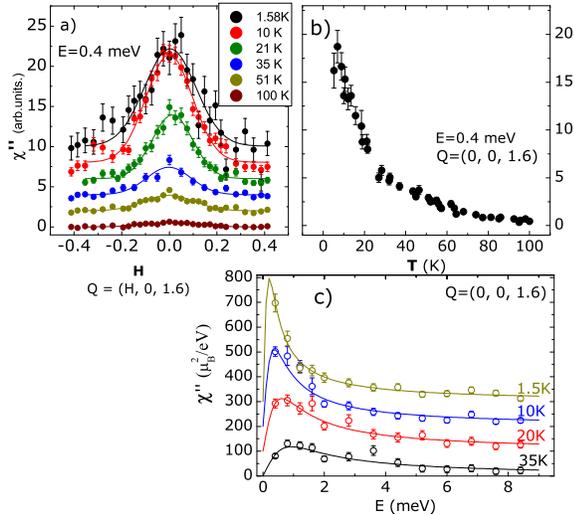}
\end{center}
\caption{(Color online) Magnetic scattering in \csreda\ taken
around the FM wave vector (0,0,1.6) on the 4F spectrometer. (a):
Constant energy scans at different temperatures. Shifted by 2 units each. (b): Signal at
(0,0,1.6) as function of temperature. (c): Signal as function of
energy at different temperatures (shifted by 100$\mu_B^2/eV$ each).
Lines are fits to a single relaxor function (see text).
\label{fmfig1}}
\end{figure}

\section{Ferromagnetic correlations}


\subsubsection*{Low-frequency ferromagnetic fluctuation}

In addition to the incommensurate scattering there is clear evidence for truly ferromagnetic scattering in \csrea
\cite{steffens07} as well as in \csreda. In Figure \ref{fmfig1} we summarize the results of constant energy scans at an
energy transfer of 0.4~meV in \csreda . The constant-Q scans were performed at \Qfm=(0,0,1.6), which is equivalent to a
ferromagnetic zone center for two-dimensional scattering. The results of single-relaxor fits are given in Fig.\ 4. By
scanning the L-component, i.~e.\ along (0,0,L) at 10~K and at a constant energy transfer of 0.4~meV, we have verified
that there is no variation of the amplitude of the signal as function of L apart from that due to the magnetic form
factor, documenting the 2D nature of this signal.

It is evident at all temperatures that the scattering is maximum at \Qfm, and there is no indication of any further
scattering at the incommensurate wave vectors. This is consistent with the presence of the incommensurate fluctuations
in \csreda, as these are known to have a much higher characteristic energy of about 2.5~meV \cite{friedt04}. Note that
in the Fig.\ \ref{fmfig1} the signal has already been corrected for the Bose factor, yielding a quantity that is
(neglecting resolution effects) proportional to the imaginary part of the susceptibility. This reveals well the
pronounced temperature dependence.

\begin{figure}
\begin{center}
\includegraphics*[width=.97\columnwidth]{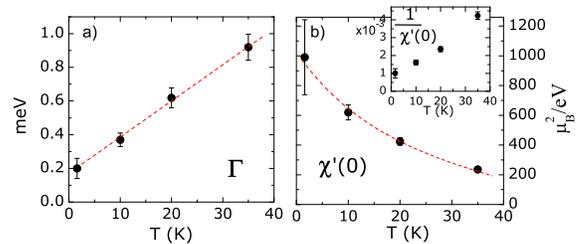}
\end{center}
\caption{(Color online) Analysis of FM scattering in \csreda. (a): Characteristic energy, as obtained from the fits
in Fig.\ \ref{fmfig1}. (b): Value of the real part of the macroscopic susceptibility, calibrated to absolute units,
obtained from the same fits. (Inset: inverse of the same values.)
\label{fmfig2}}
\end{figure}

In part (c) of Fig.\ \ref{fmfig1} we show fits with a single-relaxor function (\ref{eqrelaxor}). This function provides
a good description of the signal and allows one to extract the characteristic energy of the signal and the real part of
the susceptibility at zero frequency, which corresponds to the macroscopic susceptibility. Figure \ref{fmfig2} contains
the results for these parameters, which are both strongly temperature dependent: the characteristic energy reaches
values as low as 0.2~meV at low temperatures, which is an order of magnitude lower than the values found for the
incommensurate scattering, and which explains that this signal has not been observed in the previous studies
\cite{friedt04} which focused on a higher energy range. The susceptibility reaches very high values at low
temperatures, but remains finite. In this context, let us mention that the unit $\mu_B^2/eV$ per Ruthenium atom
corresponds to $3.23\cdot10^{-5}\,\textrm{emu}\cdot\textrm{mol}^{-1}$ in cgs units and (for the given volume of the
unit cell) to $7.102\cdot10^{-6}$ in SI. The obtained values agree with the bulk measurement in view of the
uncertainties related to the calibration process. The value at low temperatures is less exact because the
characteristic energy is so low that the maximum as function of energy transfer could not well be captured by the
neutron measurement (Fig.\ \ref{fmfig1}c). The good agreement with the macroscopic susceptibility shows that these
fluctuations are indeed the relevant ones for the observed magnetic properties, in particular for the metamagnetism.

When approaching a transition to an ordered state from above the critical temperature, it is expected that the
characteristic energy of the fluctuations approaches zero and that the susceptibility diverges such that $\chi^{-1}$
vanishes. The evolution of $\Gamma$ and $\chi$ qualitatively agrees with a transition to a ferromagnetic state.
However, the transition is not reached at finite temperatures, as the extrapolation of $\Gamma$ and $\chi^{-1}$ would
reach zero about 10~K below zero temperature. At the temperature of 50~mK there is no indication of magnetic order in
\csreda \ (nor in Ca$_{1.5}$Sr$_{0.5}$RuO$_4$ \cite{friedt04}) and the amplitude of the fluctuations is consistent with
the just discussed temperature evolution. This behavior can be compared to that of the incommensurate scattering in
\sro , which also indicates a magnetic transition that is not reached at finite temperature \cite{braden02sro}. In both
cases the blocking of the transition at low temperature may be connected with a reduction in the electrical
resistivity. In \csreda\ the suppression of the phase transition at low temperature may be further related to the
thermal expansion anomaly which although being much weaker compared to that in \csrea \ is still sizeable in \csreda
\cite{kriener05}.

%
%
%
%

\begin{figure}
\begin{center}
\includegraphics*[angle=270,width=.98\columnwidth]{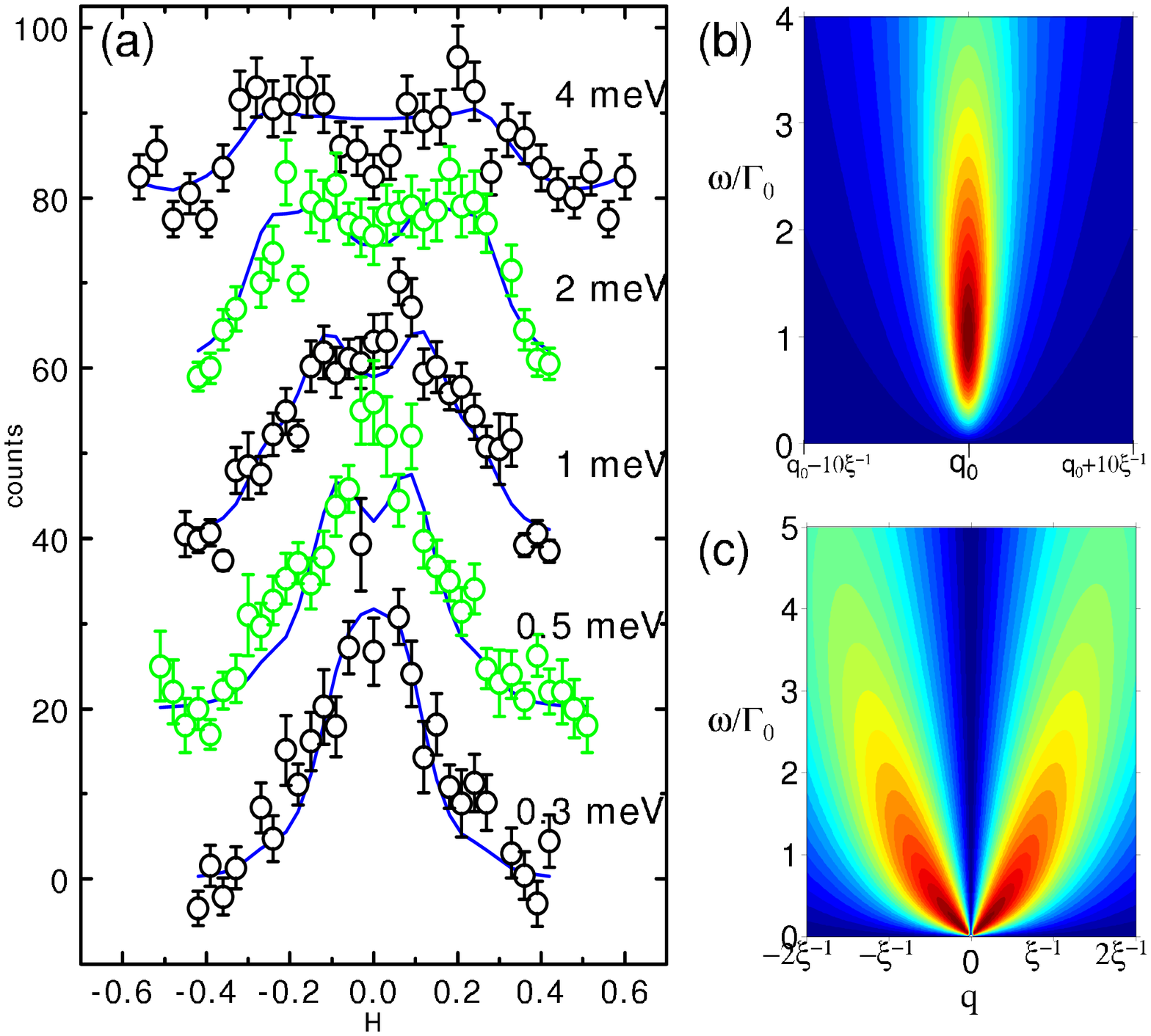}
\end{center}
\caption{(Color online) Magnetic scattering in \csreda\ (T=0.05~K, shifted by 20 counts each).
The model used to fit the data in (a) consists of a ferromagnetic
contribution and two incommensurate AFM-like excitations at \Qice\
and \Qicz. These components are given by equations
(\ref{eqrelaxor})-(\ref{eqanti}) and are displayed separately in
part (b) and (c). The FM one has to be modified to account for the
measured finite energy at the zone center. Data taken on the IN14
spectrometer.\label{fitfig1}}
\end{figure}

\subsubsection*{Model for incommensurate and ferromagnetic fluctuations in \csreda\ }

When using the corresponding equations (\ref{eqrelaxor}) and (\ref{eqferro}), the fit of the ferromagnetic component is
not satisfactory, contrary to the description of the incommensurate fluctuations. This is due to the fact that it
cannot account for the observed finite energy of the excitation in the zone centre of about $\Gamma_0$=0.2~meV as
discussed above (see for instance Fig.\ \ref{fmfig1} and \ref{fmfig2}), and the resulting maximum at H=0 in the scans
at low energy transfers in Figure \ref{fitfig1}. According to (\ref{eqrelaxor}) and (\ref{eqferro}), $\chi''(\omega)$
is zero for $q=0$, reflecting the conservation of total magnetization in the simple underlying model. In the presence
of spin-orbit coupling, for instance, this is no longer required, though it is not evident how to modify equation
(\ref{eqferro}). Another physical reason for the finite energy at q=0 may be that close to the magnetic instability,
which is actually three dimensional, the assumption of purely two dimensional fluctuations is no longer strictly
correct, or in other words, the correlation length along $c$ is no longer zero. This latter effect may be
straightforwardly included in (\ref{eqferro}) and produces a better description of the data (lines in Fig.\
\ref{fitfig1}). The value $\xi^c_\text{fm}\simeq 2\text{\AA}$ is only a phenomenological parameter, and the more
significant reason for the finite $\Gamma$ at the zone centre is likely the spin-orbit interaction. In this context we
mention that a similar effect has been observed in the paramagnetic states of UGe$_2$~\cite{huxley03}, a strongly
anisotropic Ising ferromagnetic metal, in UPt$_3$ \cite{upt3} and in MnP~\cite{yamada87}.

With the modification of a finite $\Gamma$ at the zone centre, the overall description of the entire data set taken at
T=1.5~K is satisfactory for \csreda . For the incommensurate parts, $\Gamma_0$ is 2.5$\pm$0.2~meV and
$\xi$=9.5$\pm$0.5~\Ang. For the ferromagnetic component, the energy scale is much lower, as already discussed,
$\Gamma_0$=0.34$\pm$0.05~meV, but the in-plane correlation length is very short, $\xi^{ab}_\text{fm}$=4.2$\pm$0.3~\Ang.


\begin{figure}
\begin{center}
\includegraphics*[width=.95\columnwidth]{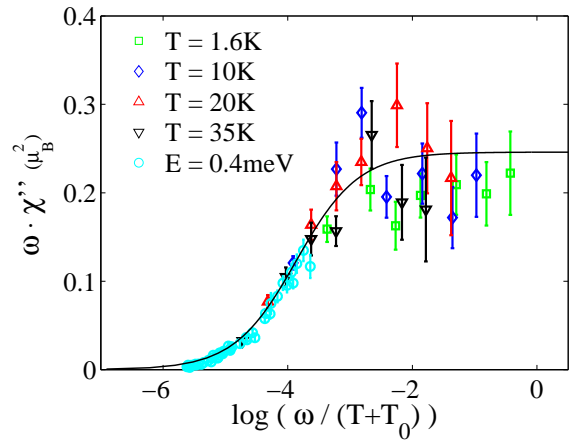}
\end{center}
\caption{(Color online) Ferromagnetic fluctuations in \csreda: different data sets taken at constant Temperature or energy
transfer (same as in Figure \ref{fmfig1}). The quantity $\omega\chi''$ only depends on $\omega/(T+\Theta)$. The line is the
function $f$ for the parameter values discussed in the text.
\label{scalfig}}
\end{figure}

The data in Fig.\ \ref{fmfig2} can be described by a simple
temperature variation of $\chi$ and $\Gamma_0$ including an offset
temperature,
\begin{equation}
\chi(T) = \frac{C}{T+\Theta}\;\;\text{and}\;\;\Gamma_0(T)= G \cdot (T+\Theta)
\label{eqtemp1}
\end{equation}

with constants $C$ and $D$. When combining these expressions with the energy spectrum (\ref{eqrelaxor}) one obtains
$\chi''(\omega,T) = C \frac{ \omega G }{\omega^2 + G^2(T+\Theta)^2}$, so it follows
\begin{equation}
\omega\cdot\chi''(\omega,T) = f\left(\frac{\omega}{T+\Theta}\right)
\end{equation}
with the function $f(x) = C\frac{x^2}{x^2+G^2}$. When plotting $\omega\chi''$ against $\omega/(T+\Theta)$, the data
should thus fall on a single curve. In Figure \ref{scalfig} this is performed for the ferromagnetic fluctuations in
\csreda\ (data from Fig.\ \ref{fmfig1} (b) and (c)), which thereby are described as function of temperature and energy.
Note however, that although resembling the Curie-Weiss susceptibility of an antiferromagnet, the meaning of the
temperature offset $\Theta$ is quite different in equation (6), where the susceptibility at the magnetic instability is
treated, whereas the antiferromagnetic Curie-Weiss law describes the macroscopic (ferromagnetic) susceptibility for an
antiferromagnetic instability.

The  analysis in Fig.\ \ref{fmfig2} allows one to obtain the values for the constants $C$, $G$ and $\Theta$. We use
$G_{FM}= 0.021 \text{meV}/K$ and $\Theta_{FM} = 10 \text{K}$. In the Curie-Weiss law, the constant $C$ is related to
the magnetic moment, $C=\frac{n\mu_0\mu_\text{eff}^2}{3k_B}$ ($n$ being the number of magnetic moments per volume).
Assuming a free local spin $\frac{1}{2}$ per Ruthenium atom, one calculates $C=11729\mu_B^2/\text{eV}\cdot\text{K}$
comparable to the value of about 11400 $\mu_B^2/\text{eV}\cdot\text{K}$ found in Fig.\ \ref{fmfig2}. This analysis
further illustrates the strength of the magnetic scattering in \csreda .

For the incommensurate fluctuations, the analysis\cite{friedt04} provides $G_{IC}= 0.1 \text{meV}/K$ and $\Theta_{IC} =
25 \text{K}$. $C_{IC}$ is, to an estimated accuracy of about 10\%, 12000 $\mu_B^2/\text{eV}\cdot\text{K}$, so
practically the same as $C_{FM}$, again documenting the strength of the magnetic correlations.

\begin{figure}
\begin{center}
\includegraphics*[width=.99\columnwidth]{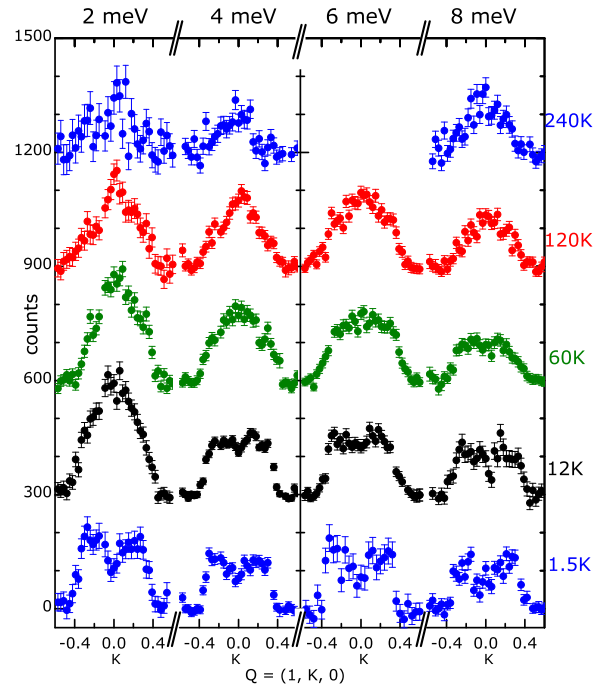}
\end{center}
\caption{(Color online) Magnetic scattering in \csreda \ as a
function of temperature and energy (background subtracted,
shifted by 300 counts each). All constant energy scans
along $Q$=(1,K,0) were performed on the thermal triple-axis
spectrometer 1T. The signal changes its shape qualitatively due to
the interplay of ferromagnetic and incommensurate contributions.
At high energy and low temperature the incommensurate signals
dominate, whereas at low energy and high temperature the centered
peak in these scans arises from the ferromagnetic
component.}\label{1tfig}
\end{figure}

With this parameter set it is, furthermore, possible to describe the interplay of ferromagnetic and incommensurate
fluctuation as function of temperature. $\Theta$ indicates how far the system at T=0 is still away from the
hypothetical magnetic instability and the divergence of the fluctuations. As $\Theta$ is larger for the incommensurate
correlations, they are less temperature dependent in the low temperature range. Experimentally, the spin fluctuations
(though broadened and with higher background) can be observed up to room temperature, because the Bose thermal factor
(eq.\ (\ref{eqsigma})) roughly compensates the decreasing magnitude of $\chi$, see Fig. \ref{1tfig}. The basic
qualitative consequence of the temperature variation (\ref{eqtemp1}) is that the balance between the incommensurate and
ferromagnetic parts changes considerably depending on the temperature and on the energy transfer of the scan. For the
given parameters and the relaxor spectrum (\ref{eqrelaxor}) it follows, for instance, that
$\chi''_\text{FM}(\omega,T)>\chi''_\text{IC}(\omega,T)$ under the condition $T>25\frac{\text{K}}{\text{meV}}\cdot
\hbar\omega - 30\text{K}$, i.~e.\ at high temperatures and/or low energy transfer one predominantly measures the
ferromagnetic fluctuations. This effect is well confirmed for x=0.62 by a series of scans at various temperatures up to
240~K and energy transfers up to 8~meV, see Fig. \ref{1tfig}. Qualitatively, the same effect is also observed for x=0.2
at higher temperatures, see Fig. 8 and below.

\subsubsection*{Suppression of ferromagnetic fluctuations at x=0.2}

\begin{figure}
\begin{center}
\includegraphics*[width=.99\columnwidth]{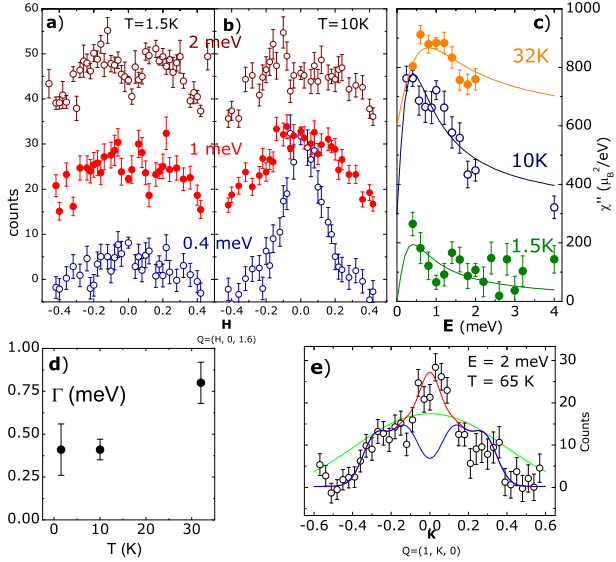}
\end{center}
\caption{(Color online) FM scattering in \csrea. (a) and (b):
Scans across \Qfm\ at different energy transfers and temperatures.
(c): Energy dependence at \Qfm\ and fits with a relaxor function
(shifted by 20 counts and 300$\mu_B^2/eV$, respectively).
(d): Characteristic energy $\Gamma$ of the signal at \Qfm.  (e): Additional ferromagnetic contribution at
T=65~K: The blue line is the fit to the corresponding scan at
T=2~K. For demonstration purposes, the green curve shows a
Gaussian broadening of the blue curve, which can produce a maximum
at K=0, but only for parameters which fail to describe the full
scan. The red line is the sum of the blue T=2~K curve and a
Gaussian peak at K=0. \label{fmfig3}}
\end{figure}

Also  \csrea \ exhibits ferromagnetic excitations, but in contrast to \csreda, they are almost entirely suppressed at
low temperatures. Figure \ref{fmfig3} summarizes the corresponding scans at 1.5 and 10~K. One might argue whether the
central intensity can arise from the overlap of the incommensurate contributions just at the ferromagnetic wave vector.
The data taken at higher temperatures and higher energy transfers unambiguously exclude this picture. Figure
\ref{fmfig3}(e) shows data taken at T=65~K for x=0.2. For comparison with low temperature, the blue line shows the
shape of the signal at T=2~K. It is impossible to ascribe the shape of the high temperature data only to a broadening;
even for less sharp central maxima, a broadening would need to be extreme to account for a maximum at K=0 (green line).
Instead, it can be very easily reproduced by simply adding an additional signal at K=0, as the one in Fig.\
\ref{fmfig1}(a) to the low temperature data. The incommensurate part of the scan is thus essentially unchanged, and the
difference between high and low temperature has to be ascribed entirely to the appearance of an additional
ferromagnetic component. Therefore, also  the susceptibility analysis and its temperature dependence has to be
interpreted as due to changes in the intrinsically ferromagnetic correlations and not to changes of the
antiferromagnetic ones.

The temperature dependence of the amplitude of the ferromagnetic signal is in good quantitative agreement with the
macroscopic susceptibility, as has already been discussed in Ref.\ \onlinecite{steffens07}. The characteristic energies
$\Gamma$ are, at the higher temperatures, of the same order as in \csreda. For x=0.2, however, it is obvious that the
relatively simple picture of approaching a magnetic instability cannot be maintained till low temperature. This is in
accordance with other anomalous effects that take place in this temperature region, in particular the remarkable
anomalous structural evolution\cite{kriener05}. A deformation of the lattice and the environment of the Ruthenium ions
occurs, that likely couples to its electronic configuration, thereby suppressing the ferromagnetic instability. It is
worthwhile to note that these effects are also present in \csreda, though to a much weaker extent\cite{kriener05}. The
close inspection of ${1\over {\chi'(0)}}$ shown in the inset of Fig. 4b) suggest that also in \csreda \ the emergence
of the ferromagnetic instability is damped by the same mechanism, but to a smaller extent.

\section{Magnetic field effect}

\begin{figure}
\begin{center}
\includegraphics*[width=.99\columnwidth]{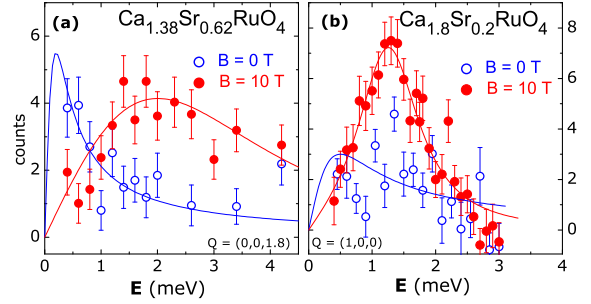}
\end{center}
\caption{(Color online) Effect of a magnetic field on the
ferromagnetic fluctuations in \csreda\ (a) and \csrea\ (b).
Temperature is 1.5~K and background has been subtracted in both
cases (but the scales are not normalized to each other). Lines
correspond to a single relaxor (B=0, with the characteristic
energies discussed in the text) and to Lorentzian functions
(B=10~T), respectively. Data were taken on the IN12 and Panda
spectrometers.\label{magfig1}}
\end{figure}

At Sr-concentrations lower than x=0.5, \csrx\ shows a metamagnetic transition\cite{nakatsuji99jlt} which manifests
itself as  a steep nonlinear increase of magnetization as a function of the external magnetic field. The metamagnetic
transition in \csrea\ has been well characterized by a number of different techniques (see for instance Refs.\
\onlinecite{nakatsuji03,balicas05,kriener05,baier07}). In particular, the inelastic neutron scattering study
\cite{steffens07} has revealed the appearance of an excitation mode that resembles a magnon in a conventional
ferromagnet. This proves that  -- whatever are its microscopic mechanism and the thermodynamic details -- a substantial
ferromagnetic interaction is induced at the metamagnetic transition.

\begin{figure}
\begin{center}
\includegraphics*[width=.99\columnwidth]{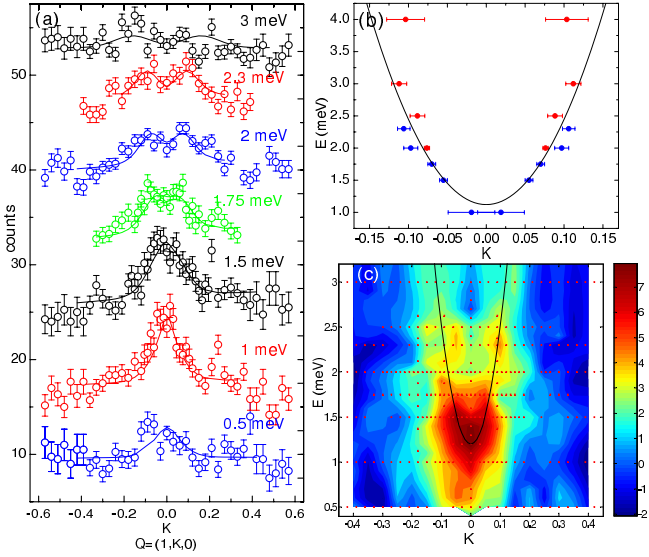}
\end{center}
\caption{(Color online) Magnetic excitations in \csrea\ at B=10~T
(B$\parallel$c and T=1.5~K).  (a): Transverse scans through
(1,0,0) at different energy transfers (shifted by 7.5 counts each).
The lines are fits to the
model described in the text. (b): Dispersion of the magnon,
extracted from fits to the scans in (a) with symmetric Gaussian
peaks. Red data points are taken from Ref.\
\onlinecite{steffens07}. (c): Colour plot of the data in (a)
(background subtracted). \label{magfig2}}
\end{figure}

\csreda\ does not show a clear metamagnetic transition, but its high susceptibility still resembles the peak at the
transition of the samples which are metamagnetic. The effect of the magnetic field on the zone-center fluctuation is
compared for \csrea\ and \csreda\ in Figure \ref{magfig1}. In both cases the zero-field response (which is very weak
for \csrea) is shifted to higher energies. The spectrum is then qualitatively different and a description using the
relaxor function (\ref{eqrelaxor}) is no longer possible.

The  energy $g\mu_BB$, which is the Zeeman energy of an electron in the magnetic field, amount to about 1.15~meV for
B=10~T. The spectral form at 10~T can indeed well be described by a Lorentzian function
$\Gamma/((\omega-\omega_0)^2+\Gamma^2)-\Gamma/((\omega+\omega_0)^2+\Gamma^2)$ with $\omega_0$=1.15~meV, the broadening,
though, being extreme (heavily overdamped) in the case of \csreda. A quantitative comparison of the widths is not
straightforward because the different crystal orientation in both cases means that the experimental resolution and the
averaging of the susceptibility components is not the same. The shift of spectral weight to higher energy is
qualitatively the same for both concentrations. Not knowing if the excitation in \csreda\ disperses in the same way as
in \csrea, it is likely that the spin correlations are weaker, as is consistent with the larger width, and in
accordance with the lower magnetic moment at 10~T compared to \csrea\ (0.4\muB\ vs.\ 0.6\muB). The ratio of the
susceptibilities obtained by Kramers-Kronig analysis $\chi'=\int \chi''(\omega)/\omega\,d\omega$ from the data in Fig.\
\ref{magfig1}a ($\chi_{0T}/\chi_{10T}\simeq 1.8$) is consistent with the ratio of the susceptibilities from the
macroscopic measurement. In \csrea, in addition, taking into account the data at intermediate fields\cite{steffens07},
the corresponding analysis perfectly describes the susceptibility maximum at the metamagnetic transition near 3~T.

With the high-field data obtained for x=0.2 it is also possible to provide a more detailed characterization of the
magnon mode at B=10~T. Because it rapidly broadens and weakens with increasing energy, it can best be studied between 1
and 3~meV; Fig.\ \ref{magfig2}a displays data taken in this energy range. The positions of the maxima fall on a
quadratic dispersion curve with a gap of 1.15~meV as displayed in Fig.\ \ref{magfig2}b, i.~e.\
\begin{equation}
\hbar\omega_q = \Delta + Dq^2
\label{eqmag1}
\end{equation}
with $ D = 47\pm 5 \,\text{meV}\cdot\text{\AA}^2 $.

It is,  however, evident from the scans and from the colour plot of the intensity in Fig.\ \ref{magfig2}c that there is
significant broadening of the magnon mode in $q$ and $\omega$. This broadening cannot be explained by the resolution of
the spectrometer, but is intrinsic. We have found that it is possible to describe the data with the usual Lorentzian
energy spectrum,
\begin{equation}
\chi''(q,\omega) \propto \frac{\Gamma_q}{(\omega-\omega_q)^2+\Gamma_q} - \frac{\Gamma_q}{(\omega+\omega_q)^2+\Gamma_q^2}
\label{eqmag2}
\end{equation}
where $\omega_q$ follows equation (\ref{eqmag1}). It is necessary, though, to include an additional parameter that
accounts for the broadening and the decrease of intensity towards higher energy -- this decrease might be due to the
approach to the Stoner continuum, where the intensity of the magnon is expected to disappear\cite{lovesey,moriya}.
Allowing $\Gamma_q$ to vary as $\Gamma_q=\Gamma_0+c\cdot q$, one introduces an additional parameter $c$ that contains a
length scale and that can be considered as modeling an effective finite spatial correlation. With this model, it is
well possible to perform a global fit to all the data that have been taken, including the resolution function of the
spectrometer. The results are $\Delta =1.16\pm0.03\,\text{meV}$, $\Gamma_0=0.55\pm0.04\,\text{meV}$,
$c=5.6\pm0.5\,\text{meV\AA}$ and $D=34\pm2\,\text{meV\AA}^2$. Note that $\Delta$, although unconstrained, corresponds
well to $g\mu_BB=1.15\,\text{meV}$ which means that anisotropy terms are either unimportant or effectively average out,
and that $D$ is different from the value given above because the new parameter $c$ shifts the maxima in the constant
energy scans with respect to the curve defined by  (\ref{eqmag1}).


\section{Conclusions}

The comprehensive INS studies provide a detailed description of the magnetic correlations in \csrx. Different types of
magnetic fluctuations are identified: we can separate a ferromagnetic signal and different features at incommensurate
Q-vectors : one on the diagonal of the Brillouin zone at \Qab=(0.3,0.3,0) and a broader and stronger contribution on
the \astar/\bstar axis. This latter one has an internal structure that can be well described by assuming two
overlapping contributions from \Qice=(0.11,0,0) and \Qicz=(0.26,0,0) and equivalent positions. Within the accuracy of
the measurement -- limited primarily by the large overlap of the signals -- there is no significant difference in the
Q-positions of these contributions for the different values of the Sr-concentration x=0.2 and x=0.62. The signal at
\Qab\ can be associated with the incommensurate signal in \sro\ \cite{sidis99}. As its origin is well understood
arising from nesting of the $\alpha$ and $\beta$ Fermi surface, the presence of this signal at the same position
indicates that these sheets of the Fermi surface, and thus also the occupation of the Ruthenium d$_{xz}$ and d$_{yz}$
orbitals are only little changed in \csrx\ (x=0.2/0.62) with respect to \sro \cite{shimoyamada09}. The origin, i.~e.\
the relevant sections of the Fermi surface, of the signals at \Qice/\Qicz\ is not yet precisely identified, but for
several reasons they are most likely related to the $\gamma$ sheet of the Fermi surface. ARPES measurements and
band-structure calculations clearly identify the $\gamma$ band as the highly renormalized one associated with the
heavy-mass electronic behavior and with the high susceptibility, but the detailed structure of the $\gamma$ sheet
awaits for further studies.

There seems to be no  significant change of the different incommensurate components for x$=$0.2 and x$=$0.62, which is
thus presumably the case for the whole range 0.18$\le$x$\le$1.5. Though the temperature dependence of $\chi$ and
$\Gamma$ of these fluctuations\cite{friedt04} indicates that the system approaches a magnetic instability at
incommensurate ordering vectors, the system can obviously be considered as still sufficiently far away and not directly
in the critical region. The structural and other variations in this range of x do not very sensitively couple to this
part of the magnetic correlations. The rotational structural distortion, however, is apparently very important and
causes the significant difference to \sro, where no excitations are observed at \Qice\ or \Qicz. It is remarkable that
in the bilayer material \srdzs\ very similar excitations as at \Qice\ and \Qicz\ have been
observed\cite{capogna03,ramos07}. In view of the similar \srdzs\ crystal structure, which also exhibits the rotational
distortion\cite{shaked00}, this appears consistent. These two layered ruthenates and their metamagnetic transitions
appear to be very similar to each other.

The incommensurate fluctuations at \Qice/\Qicz\ have a characteristic energy of about 2.7~meV, while the characteristic
energy of the excitations at the zone center is only 0.4~meV at T=10~K in \csrea. In \csreda\ this latter value further
decreases to 0.2~meV at T=1.5~K and the amplitude increases, consistent with the picture of \csreda\ approaching a
ferromagnetic instability, though not reaching it at finite temperatures. In \csrea, the ferromagnetic part of the
response is strongly suppressed at low temperature. The values of the susceptibility related to the ferromagnetic
signal are, concerning their absolute values as well as their variations with temperature and magnetic field, in
perfect agreement with the macroscopically determined susceptibilities, proving that this INS signal reflects the
magnetic correlations that determine the macroscopic physical properties.

The application of an external magnetic field at low temperature suppresses the incommensurate part of the response. In
\csrea, a substantial ferromagnetic component reappears, reflecting the metamagnetic transition. At high field, the
spectral weight of the ferromagnetic response is shifted towards higher energy for x=0.2 and 0.62, opening a gap that
roughly corresponds to the Zeeman energy of an electron in the magnetic field. In \csrea\ a dispersive excitation mode,
corresponding to a magnon in a ferromagnet, is observed, which is well defined near the zone center and low energies,
and which significantly broadens at energies above 3~meV.

Although only two concentrations have been studied here, x=0.2 and 0.62, the results are most likely of relevance for
the whole range 0.2$\le$x$\le$1.5 of the series \csrx. In this region of the phase diagram, the materials are
paramagnetic and metallic at all temperatures. The samples with x=0.2 and 0.62 have, though, to be regarded as quite
different concerning their physical properties, as there is a second-order structural phase transition (associated with
\ruo\ octahedra tilting) at x=0.5; \csrx\ seems approaches the ferromagnetic instability for x decreasing towards 0.5,
while a strongly reduced susceptibility and the metamagnetic transition are observed at x$<$0.5. It is thus important
to realize that the measured magnetic correlations can be well described in a common model in which the change of only
few parameters leads to the large differences between the nearly ferro- and nearly antiferromagnetic compounds.

The ferromagnetic correlations in \csrx\ depend sensitively on the Sr-concentration x and seem to tune the different
magnetic properties with the doping. The ferromagnetic instability in general arises from the high density of states
close to the Fermi surface of the $\gamma$ sheet, which has been reveals in several ARPES studies\cite{
wang04,shimoyamada09,neupane09}. Due to this high density of states the system becomes very sensitive upon structural
changes, as it is best documented in the thermal expansion anomalies \cite{kriener05,baier07}.

In conclusion, the comprehensive INS experiments clearly reveal the image of competing magnetic instabilities in \csrx
. The whole spectrum of magnetic correlations is well described in the presented phenomenological model with
ferromagnetic and incommensurate antiferromagnetic contributions. In this competition, it is mainly the variation of
the ferromagnetic component as function of Sr-concentration x, temperature and magnetic field, that governs the
physical properties.

\paragraph*{Acknowledgments.} This work was supported by the Deutsche Forschungsgemeinschaft through SFB 608.

\end{document}